\documentclass[runningheads]{llncs}

\usepackage{graphicx}
\usepackage{amsmath, amssymb}
\usepackage{booktabs, subcaption}
\usepackage[utf8]{inputenc}
\usepackage[T1]{fontenc}
\usepackage{listings}
\usepackage{placeins, float}
\usepackage{minted}
\usepackage{alltt}

\usepackage[hidelinks]{hyperref}
\usepackage{url}

\hypersetup{
  pdftitle={Property-Guided Cyber-Physical Reduction and Surrogation for Safety Analysis in Robotic Vehicles},
  pdfauthor={Anonymous},
  pdfkeywords={}
}

\makeatletter
\AtBeginDocument{%
  \def\doi#1{\url{https://doi.org/#1}}}
\makeatother

\begin{document}

\title{Property-Guided Cyber-Physical Reduction and Surrogation for Safety Analysis in Robotic Vehicles}
\titlerunning{Property-Guided Cyber-Physical Reduction}

\author{Nazmus Shakib Sayom\inst{1} \and Luis A. Garcia\inst{1}}
\authorrunning{N. S. Sayom and L. A. Garcia}
\institute{University of Utah, Salt Lake City, UT, USA \\ \email{\{sayom.shakib,la.garcia\}@utah.edu}}

\maketitle

\begin{abstract}
We propose a methodology for falsifying safety properties in robotic vehicle systems through property-guided reduction and surrogate execution. By isolating only the control logic and physical dynamics relevant to a given specification, we construct lightweight surrogate models that preserve property-relevant behaviors while eliminating unrelated system complexity. This enables scalable falsification via trace analysis and temporal logic oracles.

We demonstrate the approach on a drone control system containing a known safety flaw. The surrogate replicates failure conditions at a fraction of the simulation cost, and a property-guided fuzzer efficiently discovers semantic violations. Our results suggest that controller reduction, when coupled with logic-aware test generation, provides a practical and scalable path toward semantic verification of cyber-physical systems.

\end{abstract}
\section{Introduction}

Modern robotic vehicle systems face an inherent verification challenge: they combine discrete control logic with continuous physical dynamics~\cite{Frazzoli2000}, forming hybrid systems that demand domain-specific formal methods and analysis techniques to manage their complexity~\cite{Seshia2017}. Traditional verification approaches either rely on exhaustive simulations, which quickly become infeasible as system complexity increases due to the state explosion problem and the computational cost of high-fidelity modeling~\cite{Derler2012,Corso2022}, or on static code analysis techniques that struggle to capture the nuanced interactions between cyber and physical components~\cite{zaddach2014avatar,pgfuzz}. This fundamental challenge has led to a verification gap for safety-critical autonomous systems like drones, where subtle interactions between control decisions and physical dynamics can lead to hazardous conditions that evade detection during development.

The problem is compounded by how current verification methodologies treat cyber-physical systems as monolithic entities, despite their inherently modular structure. While these approaches often construct targeted mission-level scenarios to validate safety properties, they do so without exploiting the modular decomposition of the software to isolate analysis to only those components relevant to the property under verification. This leads to verification procedures that analyze components unrelated to the safety property under analysis, introducing unnecessary computational overhead. Furthermore, while recent advances like PGFUZZ~\cite{pgfuzz} and ROCAS~\cite{rocas} have made progress in detection of misconfigurations and parameter-related vulnerabilities, they are not designed to uncover deeper semantic flaws in the control logic — the kinds of issues that can cause catastrophic failures in safety-critical applications.

Recent efforts have enriched the verification landscape with techniques that address specific vulnerabilities in robotic systems. For example, policy and control-guided fuzzing frameworks such as PGFUZZ~\cite{pgfuzz} and RVFUZZER~\cite{rvfuzzer} generate inputs that stress control parameters or trigger policy violations, revealing misconfigurations and input validation flaws. CPFuzz~\cite{Shang2020CPFuzzCF} integrates signal temporal logic robustness into a coverage-guided framework, uncovering violations of formal specifications in small embedded control programs.Complementary approaches like DiscoFuzzer~\cite{Rivera2020DiscoFuzzer} detect behavioral anomalies through output discontinuity analysis, and RoboFuzz~\cite{Kim2022RoboFuzz} leverages domain-specific oracles to flag semantic violations within simulation-based testbeds. Post hoc analysis techniques such as ROCAS~\cite{rocas} bring novel capabilities in root cause localization by co-mutating both cyber parameters and environmental conditions to identify accident-triggering factors. Collectively, these approaches have significantly improved the ability to detect misconfigurations, parameter errors, and certain forms of semantic inconsistency.

However, what remains largely unaddressed is the challenge of identifying \emph{deep semantic flaws} in control logic that only surface under narrowly scoped, property-specific conditions. Existing techniques typically operate over the entire system—either through repeated full-execution simulations or by broad-brush fuzzing of high-level policies—without leveraging the internal structure of the system to localize analysis. As a result, verification remains both computationally intensive and semantically diffuse, especially when high-fidelity physical models are involved. Moreover, many current methods are tightly coupled to specific software frameworks such as ROS~\cite{rosorg} or Apollo~\cite{apolloauto_apollo}, limiting their ability to generalize across diverse cyber-physical architectures or to reason about system semantics independent of middleware implementation. Crucially, there exists no principled framework for formally reducing a cyber-physical system to the \emph{minimal executable slice} necessary to analyze a given safety or correctness property. This lack of structural and property-specific reduction limits our ability to reason efficiently about the interactions between control decisions and physical dynamics—precisely where critical bugs often reside.

To address these challenges, we introduce a verification methodology that reduces both the cyber and physical dimensions of a robotic vehicle system with respect to a given safety property. On the cyber side, we isolate the subset of control logic components that are semantically relevant to the property under analysis. These components are then composed into a property-specific surrogate system that maintains the input-output and control-flow behaviors necessary to evaluate the property, while abstracting away unrelated logic. On the physical side, we replace the full-order plant model with a reduced-order approximation that captures only the dynamics pertinent to the verification task. This reduction is designed to preserve the control responses and physical interactions relevant to the safety property, ensuring that all behaviors necessary for detecting potential violations are reta`ined. The resulting surrogate system enables targeted analysis over a minimal but semantically complete slice of the original system, improving efficiency and precision in detecting property violations.

\paragraph{Contributions.}
In this work, we present a property-guided verification methodology that reduces both the cyber and physical dimensions of robotic systems with respect to specific safety properties. Our primary contributions are as follows:

\begin{itemize}
    \item \textbf{A formalism for property-guided cyber-space reduction}, which identifies and isolates only those control logic components that are causally relevant to a given safety property through structural and dataflow analysis.
    
    \item \textbf{A host-agnostic surrogate construction framework} that enables dynamic analysis of the isolated control logic outside its original software and hardware environment, while preserving semantic equivalence with respect to the property under test.

    \item \textbf{A property-scoped physical reduction technique} that replaces full-system simulation with reduced-order models capturing only those physical dynamics that interact meaningfully with the relevant control logic.

    \item \textbf{A co-reduction workflow} that integrates cyber and physical reductions to enable efficient, property-targeted fuzzing, significantly reducing the analysis scope while retaining coverage of behaviors critical to property violations.
\end{itemize}

We demonstrate the effectiveness of our methodology through a robotic vehicle case study, illustrating how the proposed reductions facilitate targeted verification of safety properties through fuzzing the surrogate system achieved by reduction.
\section{Background}
\label{sec:background}

\subsection{Cyber-Physical Systems and Verification Challenges}
Cyber-physical systems (CPS) integrate computational processes with physical dynamics, creating hybrid systems where discrete control logic governs continuous physical behaviors~\cite{Frazzoli2000,Seshia2017}. In robotic vehicles, this manifests as control software making discrete decisions based on sensor inputs that directly affect continuous physical dynamics like motion, energy consumption, and environmental interaction. The verification of such systems presents unique challenges due to the complex interplay between cyber and physical components, where subtle bugs in control logic can lead to safety violations that only emerge under specific physical conditions~\cite{Derler2012}.

Traditional verification approaches for CPS face a fundamental scalability challenge. Exhaustive simulation-based verification becomes computationally intractable as system complexity grows, due to both the state explosion problem and the computational cost of high-fidelity physical modeling~\cite{Corso2022}. Static analysis techniques, while computationally efficient, struggle to capture the nuanced interactions between discrete control decisions and continuous physical dynamics that characterize CPS behavior~\cite{zaddach2014avatar}.

\subsection{Modularity and System Structure}
Despite being deployed as monolithic systems, modern robotic platforms like PX4~\cite{px4} and ArduPilot~\cite{ardupilot} are built from modular components that interact through well-defined interfaces. These systems typically comprise distinct modules for navigation, safety monitoring, estimation, and control, each responsible for specific aspects of system behavior~\cite{wang2022autocps}. Similarly, the physical behavior of robotic systems emerges from multiple interacting subsystems—propulsion, energy management, kinematics—each governed by distinct physical principles.

This inherent modularity presents an opportunity for targeted verification. Many safety properties depend on only a subset of the available system components. For example, a property concerning emergency parachute deployment may depend primarily on altitude estimation, battery monitoring, and emergency control logic, while being largely independent of lateral navigation or obstacle avoidance subsystems. Current verification approaches, however, treat CPS as atomic units, analyzing all components regardless of their relevance to the property under verification.

\subsection{Property-Guided Analysis}
The concept of property-guided analysis has emerged as a promising approach to address the scalability challenges in CPS verification. Rather than analyzing the entire system uniformly, property-guided techniques focus computational resources on those aspects of the system that are most relevant to the property being verified. This approach has been successfully applied in software verification through techniques like program slicing~\cite{weiser1984program}, which extracts minimal program fragments that preserve specific behavioral properties.

In the context of CPS, property-guided analysis must account for both cyber and physical dimensions. The challenge lies in identifying which control logic components and which aspects of the physical dynamics are necessary to preserve the semantics of a given safety property. This requires formal techniques for reasoning about component dependencies and interaction patterns within hybrid cyber-physical systems.

\subsection{Reduced-Order Modeling in Engineering}
Reduced-order modeling (ROM) provides a principled approach to simplifying complex dynamical systems while preserving their essential characteristics. These techniques, widely used in computational mechanics and control theory, project high-dimensional system behavior onto lower-dimensional spaces that capture the system's dominant modes~\cite{benner2015survey}. Methods such as proper orthogonal decomposition, balanced truncation, and static condensation enable efficient simulation of complex physical systems without sacrificing accuracy for the quantities of interest.

In verification contexts, reduced-order models offer a pathway to address the computational bottleneck of high-fidelity physical simulation. By focusing on only those physical dynamics that interact meaningfully with the property under analysis, ROM techniques can significantly reduce simulation time while preserving the fidelity necessary to detect property violations.

\subsection{Simulation-Based Verification}
Simulation remains a cornerstone of CPS verification, particularly for systems where mathematical modeling or formal verification is intractable~\cite{lee2017introduction}. Software-in-the-loop (SITL) simulation environments execute unmodified control software while providing simulated sensor inputs and interpreting actuator commands, enabling system-level testing without physical hardware~\cite{derler2013cyber}. This approach is fundamental to the development and verification of robotic platforms.

However, high-fidelity SITL simulations are computationally expensive. Each simulation trial requires full initialization of the software stack and detailed modeling of the physical environment and vehicle dynamics. Property verification through simulation often requires exploring large parameter spaces or validating stochastic scenarios, necessitating thousands of simulation runs. This computational burden becomes a significant bottleneck when attempting to systematically explore edge cases or validate safety properties across diverse operational conditions.
\section{Methodology}
\label{sec:methodology}
Our approach enables efficient analysis of robotic systems by reducing their cyber-physical execution to only the control logic and physical behaviors relevant to a given safety or correctness property. To illustrate the methodology, we use a representative drone scenario—modeled as a hybrid dynamical system—that captures a bug pattern inspired by a real firmware flaw discovered by PGFuzz~\cite{pgfuzz}.

\begin{figure*}[!htb]
    \centering
    \includegraphics[width=\textwidth, height=11.5cm, keepaspectratio]{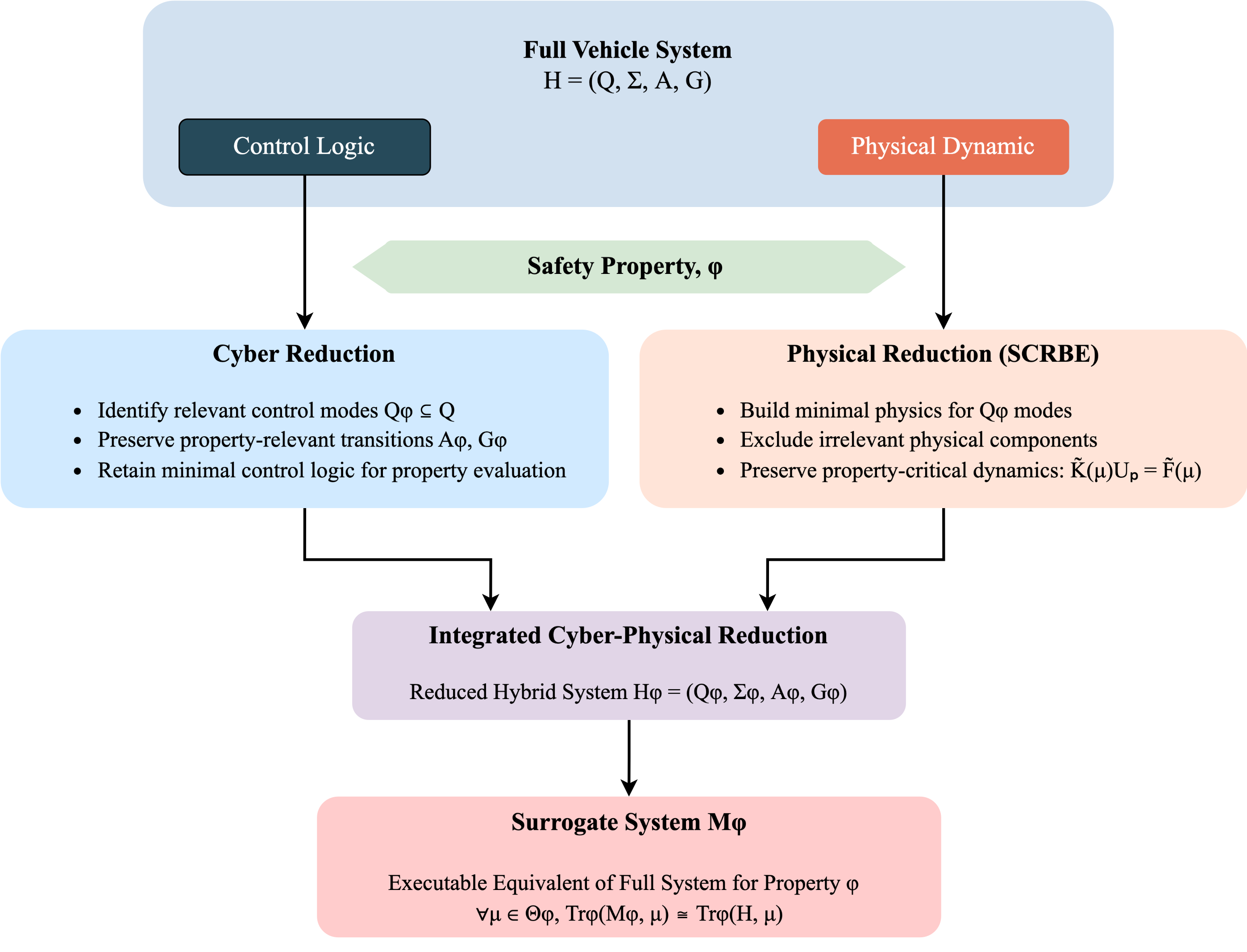}
    \caption{Overview of our cyber-physical reduction methodology for safety analysis of robotic vehicles. The approach combines control logic reduction with physical model reduction through static condensation to create an efficient, property-specific surrogate system.}
    \label{fig:methodology}
\end{figure*}

\subsection{Cyber-Physical Reduction and Structured Surrogation}

We develop a structured methodology for reducing cyber-physical systems by isolating the minimal set of components necessary for analyzing a given safety or correctness property. Our approach combines two complementary formalisms: static condensation for efficient reduction of physical component models and hybrid dynamical systems for principled pruning of control modes and transitions while preserving property-relevant behaviors.

\subsubsection{Static Condensation for Physical Model Reduction}
Our physical model reduction is grounded in the SCRBE framework of Kapteyn et al.~\cite{kapteyn2021scrbe}, which extends static condensation to modular systems with coupled physical processes. This approach enables us to create compact physical surrogate models that preserve the input-output relationships essential for property analysis.

Let the physical system be governed by a parametric variational problem:
\begin{equation}
    a(u,v;\mu) = f(v;\mu), \quad \forall v \in X(\mu),
\end{equation}
where $u(\mu)$ denotes the state of the full physical system, and $\mu$ is a high-dimensional parameter vector spanning environmental and configuration uncertainties. After discretization, we obtain:
\begin{equation}
    K(\mu) U(\mu) = F(\mu),
\end{equation}
with $K(\mu) \in \mathbb{R}^{N \times N}$ and $U(\mu) \in \mathbb{R}^{N}$. 

To restrict physical analysis to a subsystem $\pi \subseteq \mathcal{S}$—e.g., those physical components interacting with a safety-critical control path—we partition $U = [U_p, U_i]^T$ into interface and internal variables and apply static condensation:
\begin{equation}
    U_i = K_{i,i}^{-1} \left( F_i - K_{i,p} U_p \right),
\end{equation}
yielding a reduced physical system over the interface variables:
\begin{equation}
    \tilde{K}(\mu) U_p = \tilde{F}(\mu),
\end{equation}
where
\begin{equation}
    \tilde{K} = K_{p,p} - K_{p,i} K_{i,i}^{-1} K_{i,p}, \quad 
    \tilde{F} = F_p - K_{p,i} K_{i,i}^{-1} F_i.
\end{equation}

This approach allows us to create a physically reduced model that focuses computational resources on the physical aspects relevant to our property of interest.

\subsubsection{Hybrid Dynamical Systems for Control Logic Reduction}
The computational (cyber) aspects of our system are modeled using hybrid dynamical systems (HDS), providing a formal framework for identifying which control modes and transitions must be preserved in our reduced model. Following Branicky's formalism~\cite{Branicky2005}, we represent the full cyber-physical system as a tuple:

$$H = (Q, \Sigma, A, G)$$

where:
\begin{itemize}
    \item $Q$ is a finite set of discrete states or modes, each corresponding to a distinct control logic configuration
    \item $\Sigma = \{\Sigma_q\}_{q \in Q}$ is a collection of continuous dynamics, where each $\Sigma_q$ is governed by $\dot{x} = f_q(x, \mu)$ on state space $X_q \subset \mathbb{R}^n$, with $\mu \in \Theta$ representing system parameters
    \item $A = \{A_q\}_{q \in Q}$ are the autonomous jump sets, where $A_q \subset X_q$ defines conditions triggering transitions from mode $q$
    \item $G = \{G_q\}_{q \in Q}$ are the transition maps, where $G_q: A_q \to S$ determines the target state after transitions ($S = \cup_{q \in Q} X_q \times \{q\}$ being the hybrid state space)
\end{itemize}

This HDS formulation makes explicit the interrelationships between control modes and physical dynamics, enabling us to reason about which elements of the control logic are necessary to preserve the behaviors relevant to our analysis.

\subsubsection{Integrated Cyber-Physical Reduction}
For a given property $\varphi$, our methodology creates a property-specific reduced system through a two-phase process:

First, we apply the HDS formalism to identify the minimal control structure necessary for analyzing $\varphi$. This yields a reduced hybrid system $H_\varphi = (Q_\varphi, \Sigma_\varphi, A_\varphi, G_\varphi)$ constructed by:

\begin{enumerate}
    \item Identifying the subset of control modes $Q_\varphi \subseteq Q$ that are causally relevant to $\varphi$
    \item Preserving only those transitions in $A_\varphi$ and $G_\varphi$ that can occur along execution paths relevant to $\varphi$
    \item Retaining the minimal control logic necessary for property evaluation
\end{enumerate}

Second, for each mode $q \in Q_\varphi$, we apply SCRBE to reduce the physical model associated with that mode, creating efficient surrogate representations of the continuous dynamics $f_q(x, \mu)$.

The integration of these two reduction approaches yields a comprehensive cyber-physical reduced model that maintains:

$$\forall \mu \in \Theta_\varphi, \quad \text{Tr}_\varphi(H_\varphi, \mu) \cong \text{Tr}_\varphi(H, \mu)$$

where $\text{Tr}_\varphi$ extracts the execution trace elements relevant to property $\varphi$, and $\cong$ denotes behavioral equivalence with respect to those elements. 

For practical analysis, we implement $H_\varphi$ as a concrete surrogate system $M_\varphi$ that provides an efficient, executable representation while preserving semantic equivalence with respect to the property:

$$\forall \mu \in \Theta_\varphi, \quad \text{Tr}_\varphi(M_\varphi, \mu) \cong \text{Tr}_\varphi(H, \mu)$$

\noindent\textit{Scope of equivalence.} In this paper, $\text{Tr}_\varphi$ denotes the projection to the signals referenced by $\varphi$ (e.g., battery, altitude, deployment). We interpret $\cong$ as agreement in the truth valuation of $\varphi$ over the projected trace. We validate this \emph{property-scoped} equivalence empirically by checking that the surrogate and the full system yield the same satisfaction or, violation outcome for the same configurations.

This reduction framework naturally distinguishes between two complementary forms of simplification: 
\emph{cyber reduction}, which isolates the minimal subset of control modes and transitions necessary for evaluating property $\varphi$; and 
\emph{physical reduction}, which applies SCRBE to create efficient surrogate models of the physical dynamics in each retained mode. 
The former corresponds to a structural restriction of the control logic and firmware execution pipeline, while the latter yields computationally efficient physical models and a reduced parameter space for analysis.
Together, these reductions enable targeted reasoning about subsystem behavior while preserving fidelity with respect to the full system along the slice relevant to $\varphi$.

\paragraph{Example: Fault-Aware Reduction of a Parachute Deployment Pipeline.}
We consider a drone system exhibiting a safety-critical bug in its emergency parachute deployment logic. When the battery falls below a configured threshold, the emergency controller is expected to deploy the parachute unconditionally. However, due to an implementation flaw, deployment is prevented if the drone is outside a safe altitude range—specifically, if it is below a minimum safe altitude or above a maximum safe altitude—an inappropriate constraint under critical battery conditions.

We model this system using a hybrid dynamical system \( H = (Q, \Sigma, A, G) \), where:

\begin{itemize}
    \item The mode set \( Q \) consists of operational states such as \texttt{IDLE}, \texttt{TAKE\_OFF}, \texttt{GOTO}, \texttt{LAND}, and \texttt{PARACHUTE}, each corresponding to a distinct phase of the drone's mission.

    \item The collection \( \Sigma = \{\Sigma_q\}_{q \in Q} \) defines the continuous-time dynamics within each mode. These dynamics govern the evolution of physical state variables such as position, velocity, and battery level.

    \item The guard sets \( A = \{A_q\}_{q \in Q} \) define the conditions under which transitions occur between modes, based on state-dependent thresholds such as altitude, position, or remaining energy.

    \item The transition maps \( G = \{G_q\}_{q \in Q} \) determine the post-transition hybrid state, including updates to the active mode and any necessary resets or reinitializations of the continuous state. For example, transitioning to \texttt{PARACHUTE} mode may reinitialize the system's velocity state to reflect a controlled descent profile.
\end{itemize}

The full hybrid system encodes not only normal mission behavior but also soft-landing maneuvers, emergency overrides, PID-stabilized position tracking, and mode-specific force modulation. It contains significant behavioral diversity beyond the logic necessary to evaluate the property \( \varphi \), making full-scale falsification or symbolic analysis computationally expensive.

We define the safety property:
\[
    \varphi = G(\texttt{battery\_low} \wedge \texttt{airborne} \rightarrow F_{[0,\Delta]} \texttt{deployed})
\]

To analyze this property efficiently, we construct a reduced hybrid system \( H_\varphi = (Q_\varphi, \Sigma_\varphi, A_\varphi, G_\varphi) \) that preserves all behavior relevant to \( \varphi \). This reduction includes:

\begin{itemize}
    \item A restricted mode set \( Q_\varphi = \{\texttt{GOTO}, \texttt{PARACHUTE} \} \), excluding modes not involved in the fault scenario.
    
    \item A state projection \( x_\varphi = (\texttt{battery}, \texttt{altitude}) \), eliminating dimensions unrelated to the triggering and outcome of the bug.
    
    \item A reduced guard set \( A_\varphi \subseteq A \), retaining only the transition predicate responsible for triggering parachute deployment under low battery.
    
    \item A corresponding transition map \( G_\varphi \subseteq G \) that encodes the system's behavior after deployment, including any reinitialization or dynamics required for continued simulation in \( \texttt{PARACHUTE} \) mode.
\end{itemize}

This reduction enables efficient detection of the guard logic flaw without executing unrelated control paths. Furthermore, we apply physical model reduction via SCRBE to retain only the simplified descent dynamics necessary for analyzing the timing and validity of deployment behavior.

The surrogate system \( M_\varphi \) thus supports efficient falsification of \(\varphi\), preserving trace-level equivalence with the original system over the property-relevant slice:

\[
    \forall \mu \in \Theta_\varphi, \quad 
    \text{Tr}_\varphi(M_\varphi, \mu) \cong \text{Tr}_\varphi(H, \mu)
\]


\subsection{Property-Guided Falsification over the Reduced System}

Once the reduced surrogate system \( M_\varphi \) has been constructed, the next step is to explore its behavior under varied configurations and runtime conditions to identify inputs that violate the safety property \( \varphi \). To this end, we employ a structured falsification process that systematically searches over the reduced parameter space \( \Theta_\varphi \) for configurations that falsify the specification.

We define the falsification objective as discovering a configuration \( \mu \in \Theta_\varphi \) such that the resulting system trace fails to satisfy the property:
\[
\text{Tr}_\varphi(M_\varphi, \mu) \not\models \varphi
\]
where \( \text{Tr}_\varphi \) denotes the trace projected to the state and control signals relevant to \( \varphi \).

Each configuration \( \mu \) comprises both physical state variables (e.g., initial altitude, battery level) and control parameters (e.g., deployment thresholds). Rather than relying on unconstrained sampling, the search process applies domain-aware generation and mutation strategies that preserve structural constraints—such as the ordering between minimum and maximum altitude—and prioritize exploration near semantically significant regions.

Evaluation of each candidate proceeds in three stages:
\begin{enumerate}
    \item Execute the reduced system \( M_\varphi(\mu) \) to produce a time-indexed execution trace.
    \item Analyze the trace using a logical oracle corresponding to the safety property \( \varphi \).
    \item If the property is violated, log the trace and configuration; otherwise, continue exploration.
\end{enumerate}

To reduce redundancy, the falsification loop retains only non-equivalent counterexamples and discards duplicate violations. The final result is a curated collection of minimal failing inputs along with their associated traces, enabling post-hoc analysis of failure regions within \( \Theta_\varphi \).

\subsubsection{Temporal Logic Oracle}

Each trace is evaluated against the specification by a temporal logic oracle that maps input configurations to a Boolean verdict:
\[
\mathcal{O}_\varphi(\mu) =
\begin{cases}
\texttt{Satisfied}, & \text{if } \text{Tr}_\varphi(M_\varphi, \mu) \models \varphi \\
\texttt{Violated}, & \text{if } \text{Tr}_\varphi(M_\varphi, \mu) \not\models \varphi
\end{cases}
\]

The property \( \varphi \), previously defined using Signal Temporal Logic (STL), expresses that any low-battery condition during flight must lead to a successful parachute deployment within a bounded delay.

The oracle implements the formal semantics of STL to determine whether the trace satisfies \( \varphi \), treating each simulation as a deterministic signal over the relevant observable variables.

This falsification loop serves as an effective mechanism for evaluating the sufficiency of the reduced model and for uncovering subtle cyber-physical logic flaws that may only manifest under specific combinations of physical state and control configuration.

\section{Implementation}

To demonstrate the viability of our methodology, we develop a proof-of-concept system that instantiates each phase of the proposed framework. While simplified in scope, this implementation embodies the full reduction-surrogation-falsification pipeline and serves as a concrete vehicle for validating the technical approach.

\begin{figure*}[!htb]
    \centering
    \includegraphics[width=12cm]{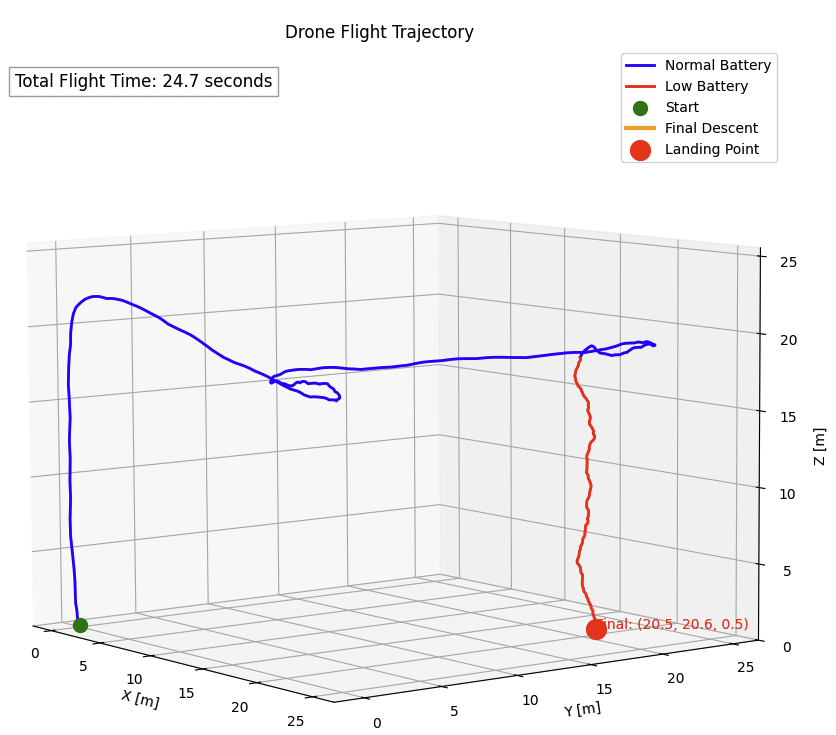}
    \caption{\textbf{Full-System Flight Trajectory.} PyBullet-based simulation trace showing 3D descent profile after battery low warning. But the drone does not deploy the parachute due to a bug in the emergency controller. Simulation took 24.7 seconds and required complete system execution.}
    \label{fig:full_trajectory}
\end{figure*}

\subsection{System Overview}

The example system models a drone flight controller responsible for emergency parachute deployment. Rather than simulating the entire PyBullet-based software stack and physical dynamics, the implementation applies structural and semantic reductions that isolate only the components relevant to evaluating a critical safety property. The resulting surrogate behaves identically with respect to the property, but is significantly more lightweight and analyzable.


\subsubsection*{Cyber-Reduction Realization}

The cyber portion of the system is distilled from a realistic multi-mode flight control stack to a minimal state machine with two control modes: waypoint navigation (\texttt{GOTO}) and emergency override (\texttt{PARACHUTE}). Intermediate states (e.g., \texttt{IDLE}, \texttt{TAKE\_OFF}, \texttt{LAND}) and unrelated control flows are omitted. The retained logic preserves mission-relevant transitions and emergency triggers, ensuring property-relevant behavior is intact.
This selection follows the formal restriction $Q_\varphi$ described in Section~\ref{sec:methodology}, guided directly by the STL property and its triggering guard conditions.

\subsubsection*{Physical Reduction Realization}

Physical dynamics are reduced to a minimal two-variable model capturing battery level and altitude—abstracting away full-body flight physics, actuator dynamics, and energy consumption models. These two variables are sufficient for evaluating deployment behavior in the presence of low-battery conditions, enabling efficient simulation and trace extraction without loss of semantic coverage.
This corresponds to the projection used by $\text{Tr}_\varphi$ and the static-condensation-based reduction in Section~\ref{sec:methodology}.

\subsubsection*{Executable Surrogate System}

The reduced cyber and physical components are integrated into a self-contained Python system that executes full traces over the reduced hybrid state space. The configuration includes parametric threshold settings (e.g., minimum/maximum deployment altitude, battery critical threshold) that control mode transitions and influence safety-critical decisions. This surrogate system offers a clean, bounded execution model suitable for rigorous automated falsification.

\subsection{Falsification Framework}

To test the resilience of the reduced system, we construct a targeted fuzzing pipeline that explores the parameter space for configurations that violate the intended safety property. The fuzzer employs constraint-preserving generators and feedback-driven mutations to systematically probe boundary conditions and uncover logic flaws.

Each run produces a complete execution trace along with a verdict from a temporal logic oracle, which analyzes the trace against a formally specified STL property. Violating configurations are logged for inspection, and visual analytics summarize the falsification landscape. The fuzzer supports continuous execution until interruption, automatically generating visualization outputs at termination.

\subsection{Overview}

This implementation demonstrates the core contribution of our framework: the ability to reduce a complex cyber-physical system to a semantically faithful, executable surrogate and to evaluate it through property-guided falsification. Despite its compactness, the system supports trace-level diagnostics, captures subtle safety violations, and enables reproducible experiments grounded in formal logic.

\section{Evaluation}
\label{sec:evaluation}


\begin{figure*}[!htb]
    \centering
    \includegraphics[width=1\textwidth]{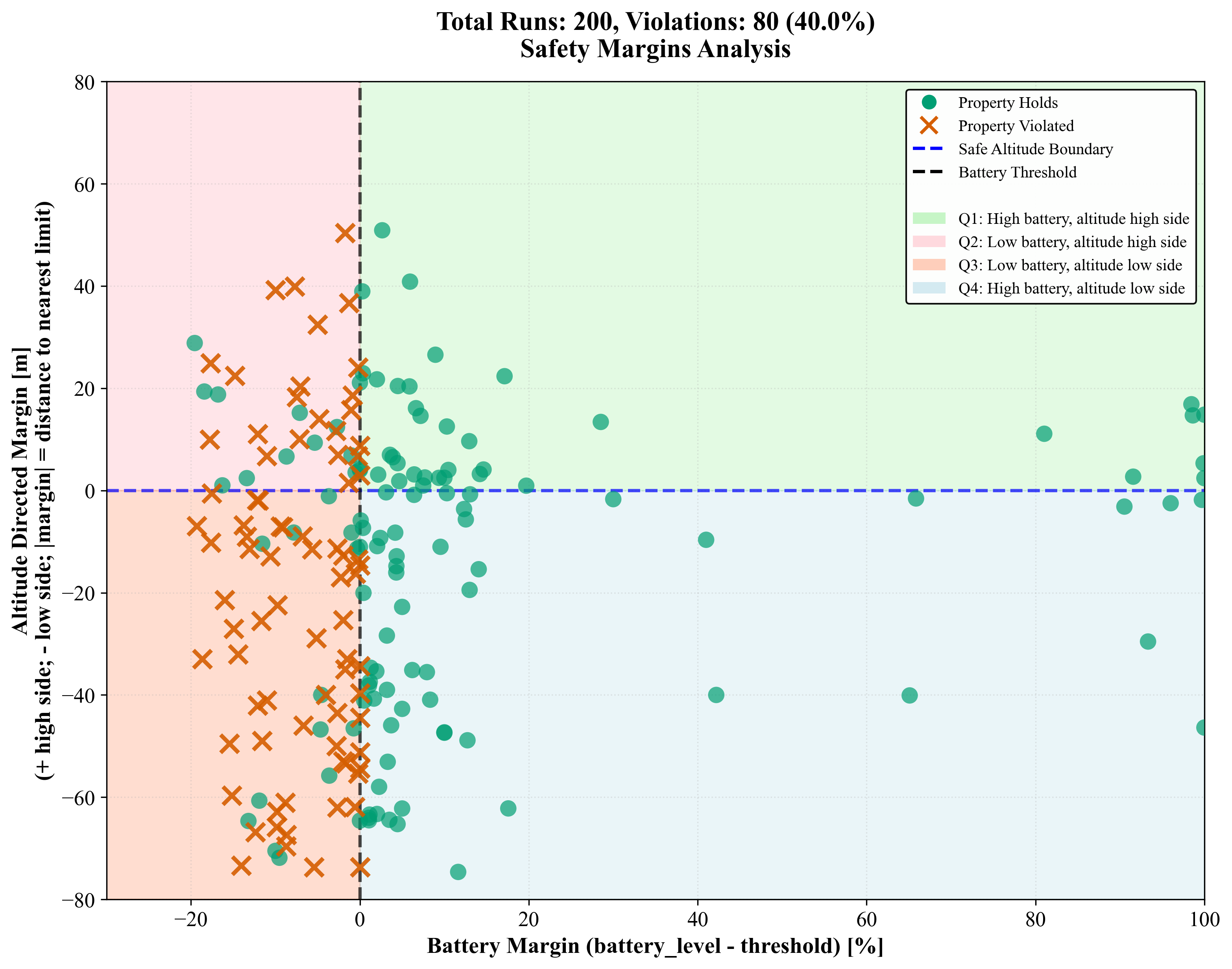}
    \caption{\textbf{Battery safety margin vs.\ altitude-directed safety margin for 200 autonomous runs.} Each point is one run: green circles indicate that the safety property was satisfied, and orange crosses indicate a violation (80/200 runs, 40\%). The battery safety margin on the $x$-axis is the observed battery level relative to the mission low-battery threshold ($M_{\mathrm{bat}} = \text{battery level} - \text{threshold}$); larger values indicate more remaining battery. The altitude-directed margin on the $y$-axis is the signed distance to the nearest allowed deployment altitude limit; positive values indicate that the vehicle is above the maximum allowed deployment altitude, negative values indicate that it is below the minimum allowed deployment altitude, and $0$ corresponds to contact with a limit. Dashed lines mark the controller's battery threshold (vertical) and safe-altitude boundary (horizontal). Shaded quadrants ($Q1$--$Q4$) separate high/low battery and high-/low-side altitude regimes. Violations cluster where the battery margin is low and the vehicle is \emph{outside} the allowed altitude band (either too low or too high), showing that the parachute deployment logic is incorrectly inhibited whenever altitude is out of range, even under critical battery conditions where deployment should be unconditional.}
    \label{fig:safety_margins}
\end{figure*}

We evaluate our methodology on a drone deployment scenario inspired by a safety-critical control flaw originally reported in PX4-based systems~\cite{pgfuzz}. Our goal is to assess whether property-guided reduction and surrogate construction enable efficient, semantically faithful falsification while retaining practical utility.

\subsection{Baseline: Full-System Simulation Cost}

We begin by attempting falsification via full PyBullet-based simulation. This process involves mission setup, realistic flight dynamics, and complete control stack execution. A single trace that reaches the faulty deployment condition takes over 24 seconds of wall-clock time (see Figure~\ref{fig:full_trajectory}).

Beyond runtime, this approach requires significant engineering effort including platform-specific instrumentation, logging infrastructure, and telemetry capture. These overheads make full-system falsification expensive and difficult to scale for proactive validation or regression testing.

\subsection{Falsification through Surrogate Execution}

Using our reduction methodology (Section~\ref{sec:methodology}), we construct a property-specific surrogate that preserves only the relevant cyber-physical behavior. Each surrogate run completes in under 500 milliseconds and produces a complete trace with STL-based property evaluation.

\vspace{0.5em}
\noindent\textbf{Example run (reduced system):}
\begin{lstlisting}[language=bash, basicstyle=\ttfamily\small, breaklines=true]
$ python run_scrbe_simulation.py --battery 10.0 --altitude 20
Configuration:
- Min deploy altitude: 60.0m
- Max deploy altitude: 80.0m
- Low battery threshold: 10.0%
Result:
Battery: 10.0% (CRITICAL)
Altitude: 20.0m
Parachute: NOT DEPLOYED
Status: BLOCKED - Critical battery but altitude out of deployment range
\end{lstlisting}

This trace mirrors the behavioral failure seen in PyBullet-based full physics simulation with full control stack, showing that the surrogate reproduces the relevant failure semantics at a fraction of the cost.

As a property-scoped conformance check for $\varphi$, we treat this agreement as evidence that the surrogate and full simulation yield the same truth valuation of $\varphi$ under the same configuration. In particular, existence of a configuration $\mu^*$ such that both traces violate $\varphi$ witnesses that $M_\varphi$ reproduces the failure semantics relevant to $\varphi$ in this scenario.

\subsection{Fuzzing Results on Reduced Model}

We implemented a feedback-guided fuzzer that perturbs both physical conditions (e.g., altitude, battery) and controller thresholds. Table~\ref{tab:fuzz_summary} summarizes the results from 20 runs on patched and non-patched controllers.

\begin{table}[htbp]
    \centering
    \caption{Fuzzing Results: Patched vs. Non-Patched Controller}
    \label{tab:fuzz_summary}
    \begin{tabular}{@{}lcc@{}}
        \toprule
         & Non-Patched & Patched \\
        \midrule
        Total Runs & 200 & 200 \\
        Unique Violations & 80 & 0 \\
        Violation Rate & 40\% & 0\% \\
        \bottomrule
    \end{tabular}
\end{table}

The unpatched system consistently violates the safety property, while the patched version satisfies it across all tested inputs—demonstrating the surrogate's utility for diagnostic validation.






\subsection{Visualizing Safety Envelope and Violation Modes}
\label{sec:violation_visualization}

We analyze how the controller behaves near its safety limits by projecting each evaluated run into a two-dimensional ``margin space'' defined by battery margin and altitude-directed margin (Fig.~\ref{fig:safety_margins}). Each point in this space corresponds to one run and is labeled according to whether the safety property was satisfied or violated. Both axes are expressed as distances to the controller's own decision boundaries: the battery axis measures remaining battery relative to the configured low-battery trigger level (positive when sufficient charge remains, negative when below threshold), and the altitude axis measures signed distance to the nearest permitted deployment altitude limit (negative when the vehicle is below the minimum deployment altitude, positive when above the maximum).

This view makes the failure mode of the emergency parachute logic explicit. The intended behavior is that, once the battery drops below its emergency threshold, deployment should proceed unconditionally. Instead, violations concentrate in regions where the battery margin is low \emph{and} the vehicle lies outside the nominal deployment altitude band. In these cases, deployment is blocked because the altitude is either too low or too high, even though that altitude check is not supposed to prevent an emergency action under critical battery conditions.

Crucially, Fig.~\ref{fig:safety_margins} is generated from the reduced surrogate model rather than the full flight stack. The surrogate reproduces the same interaction between low battery and altitude gating that we observe in the full system, and it produces the same class of safety violations (parachute withheld despite critical battery). This provides empirical evidence that the surrogate preserves the safety-relevant decision logic of the original controller, and therefore can be fuzzed in place of the full system while still revealing the true violation modes.

\subsection{Effectiveness and Scalability}

Our approach demonstrates several benefits:

\begin{itemize}
    \item \textbf{Semantic fidelity (w.r.t. $\varphi$):} The surrogate reproduces failure behavior under the same conditions as the whole system simulated in a full physics simulation.
    \item \textbf{Runtime efficiency:} Surrogate runs are over 10$\times$ faster, enabling large-scale test campaigns.
    \item \textbf{Fuzzing tractability:} Temporal falsification becomes practical over high-dimensional configuration spaces.
\end{itemize}

This evaluation confirms that our surrogate-based methodology enables efficient falsification and insight-driven debugging. It faithfully mirrors full-system bugs while reducing infrastructure burden and runtime costs.
\section{Discussion}
\label{sec:discussion}

This work establishes a principled framework for cyber-physical verification through property-guided reduction and surrogate execution. Our methodology demonstrates how formal reduction techniques can be systematically applied to isolate and analyze the minimal components necessary for property verification in robotic systems.

\noindent\textbf{Theoretical Foundation and Generalizability.} The core contribution of this work lies in establishing the formal foundation for property-guided cyber-physical reduction. By grounding cyber reduction in hybrid dynamical systems theory and physical reduction in static condensation, we provide a domain-agnostic methodology that applies broadly across CPS platforms. The framework's theoretical basis ensures that reductions preserve semantic equivalence with respect to the target property, enabling reliable verification on simplified surrogate systems.

\noindent\textbf{Methodological Innovation.} Our approach represents a significant departure from traditional monolithic verification strategies. Rather than treating cyber-physical systems as atomic units, we demonstrate how to systematically decompose them into property-relevant components while maintaining behavioral fidelity. This decomposition enables targeted analysis that focuses computational resources precisely where they are needed, addressing the fundamental scalability challenges that plague existing verification approaches.

\noindent\textbf{Practical Efficiency and Automation Potential.} The surrogate-based workflow we establish provides substantial computational advantages over full-system simulation. Our evaluation demonstrates significant runtime improvements while maintaining complete accuracy in bug detection. Importantly, the formal structure of our reduction methodology creates natural opportunities for automation—the systematic component identification and interface analysis procedures we define can serve as the foundation for automated tooling in future implementations.

\noindent\textbf{Validation and Semantic Consistency.} A critical achievement of our framework is the preservation of semantic equivalence between original systems and their reduced surrogates. Grounded in formal models, our methodology is evaluated empirically: we validate property‑level semantic preservation by reproducing the same violation conditions in the surrogate and full‑system settings. This semantic consistency is essential for practical adoption of the methodology.

\noindent\textbf{Architectural Compatibility.} Our approach leverages the modular architecture inherent in modern robotic platforms, working with well-established open-source systems like PX4 and ArduPilot. This compatibility with existing development ecosystems positions the methodology for practical adoption without requiring fundamental changes to current robotic software engineering practices.

\noindent\textbf{Evaluation Rigor.} The focused evaluation we present serves multiple purposes: it validates the theoretical soundness of our reduction techniques, demonstrates practical efficiency gains, and establishes a reproducible baseline for future comparative studies. The case study provides clear evidence that property-guided reduction can achieve substantial computational savings while maintaining verification accuracy.

Our framework establishes the theoretical and methodological foundation for a new class of property-specific verification tools. The systematic reduction techniques we present enable efficient analysis of semantic flaws in cyber-physical systems, offering a principled alternative to computationally expensive full-system approaches. The formal grounding of our methodology provides a solid foundation for future extensions and automation efforts in property-guided CPS verification. 

\noindent\textbf{Limitations and scope.} Our evaluation focuses on a single robotic platform, one representative safety property, and one bug manifestation. This scope is intentional: our primary contribution is methodological, centered on property-guided cyber-physical reduction with surrogation. The study is designed to isolate feasibility and demonstrate the end-to-end workflow. Broadening across platforms and properties requires substantial re-instrumentation and scenario curation, which is orthogonal to the core technique. Within this scoped setting, we demonstrate that the reductions preserve safety-relevant behavior while delivering substantial efficiency gains that enable practical analysis. Future work will expand the evaluation to additional platforms, properties, and bug types to further validate generalizability.
\section{Conclusion}

We have presented a property-guided methodology for cyber-physical system verification that enables semantically faithful falsification through structural reduction and surrogate construction. By isolating the minimal control and physical components relevant to a given safety property, our approach transforms complex robotic systems into lightweight, analyzable surrogates—supporting scalable, specification-driven testing without reliance on full-system simulation.

Through a proof-of-concept implementation targeting a real-world drone deployment bug, we demonstrated how this reduction enables targeted fuzzing, rapid falsification, and interpretable diagnostics with minimal engineering overhead. Our evaluation confirms that safety violations can be exposed in orders-of-magnitude less time while preserving behavioral equivalence with the full system.

This work opens the door to broader verification frameworks that are property-aware by construction. Future research will focus on automating reduction workflows, scaling to multi-property analyses, and applying these techniques to closed-source and industry-scale platforms.

\subsubsection*{Acknowledgements.} The research reported in this paper was sponsored in part by the 
National Science Foundation (NSF) under awards FMitF-2425711 and NCS-2435642 and the National Institute of Health (NIH) under award R61MH135109. The views and conclusions contained in this document are those of the authors and should not be interpreted as representing the official policies, either expressed or implied, of the funding agencies. 

\clearpage

\bibliographystyle{splncs04}
\bibliography{Sources/references}

\end{document}